# Raman Fingerprint of Pressure-Induced Phase Transitions in TiS$_3$ Nanoribbons: Implications for Thermal Measurements under Extreme Stress Conditions

*K. K. Mishra[1], T. R. Ravindran[2], Joshua O. Island[3], Eduardo Flores[4], Jose Ramon Ares[5], Carlos Sanchez[5], Isabel J. Ferrer[5], Herre S. J. van der Zant[3], Amit Pawbake[6], R. Kanawade[7], Andres Castellanos-Gomez[8] and Dattatray J. Late[9,\*]*

[1]Department of Physics and Institute for Functional Nanomaterials, University of Puerto Rico, San Juan, PR 00931, USA

[2]HBNI, Materials Science Group, Indira Gandhi Centre for Atomic Research, Kalpakkam 603102, India

[3]Kavli Institute of Nanoscience, Delft University of Technology, Lorentzweg 1, 2628 CJ Delft, Netherlands

[4]FINDER-group, Instituto de Micro y Nanotecnología, IMN-CNM, CSIC (CEI UAM+CSIC) Isaac Newton, 8, E-28760, Tres Cantos, Madrid (Spain)

[5]Materials of Interest in Renewable Energies Group (MIRE Group), Departamento de Física de Materiales, Universidad Autónoma de Madrid, UAM, 28049 Madrid, Spain

[6]J. Heyrovský Institute of Physical Chemistry, Czech Academy of Sciences, Dolejškova 2155/3, 182 23 Prague, Czech Republic

[7]Department of Physics, Savitribai Phule Pune University, Pune-411008, India

[8]Instituto Madrileño de Estudios Avanzados en Nanociencia (IMDEA Nanociencia), Campus de Cantoblanco, E-28049 Madrid, Spain

[9]Centre for Nanoscience & Nanotechnology, Amity University Maharashtra, Mumbai - Pune Expressway, Bhatan, Post-Somathne, Panvel, Maharashtra-410206, India

[\*]Corresponding author: datta099@gmail.com

**ABSTRACT**

Two-dimensional layered trichalcogenide materials have recently attracted the attention of the scientific community because of its robust mechanical, thermal properties and applications in opto and nanoelectronics devices. We report the pressure dependence of out-of plane $A_g$ Raman modes in high quality few-layers titanium trisulfide ($TiS_3$) nanoribbons grown using a direct solid-gas reaction method and infer their cross-plane thermal expansion coefficient.Both mechanical stability and thermal properties of the $TiS_3$ nanoribbons are elucidated using phonon-spectrum analyses. Raman spectroscopic studies at high pressure (up to 34 GPa) using a diamond anvil cell identify four prominent $A_g$ Raman bands; a band at 557 cm$^{-1}$ softens under compression, and others at 175, 300, and 370 cm$^{-1}$ show normal hardening. Anomalies in phonon mode frequencies and excessive broadening in line-width of the soft phonon about ~ 13 GPa are attributed to the possible onset of a reversible structural transition. A complete structural phase transition at 43 GPa is inferred from Ag soft mode frequency (557 cm$^{-1}$) versus pressure extrapolation curve, consistent with recent reported theoretical predictions. Using the experimental mode Grüneisen parameters $\gamma_i$ of Raman modes, the cross-plane thermal expansion coefficient $C_v$ of the $TiS_3$ nanoribbons at ambient phase is estimated to be $1.321 \times 10^{-6}$K$^{-1}$. The observed results are expected to be useful in calibration and performance of next generation nano-electronics and optical devices under extreme stress conditions.

# INTRODUCTION

Two dimensional (2D) van der Waals thin layer structure materials such as $MoS_2$, $WS_2$, $WSe_2$, $MoSe_2$, GaS, GaSe, h-BN, $SnSe_2$, $TiS_3$, GeTe, $MoTe_2$, $WTe_2$ and black phosphorous have attracted significant research interest because of their extraordinary physical, optical and electronic properties[1-7]. These materials are found to be suitable for a wide range of applications such as gas sensors[3,8,9], photo detectors[10-15], transistors[12,16-20], solar cells[21,22], energy storage devices[23] and field emitters[13,24-27]. The central feature of this 2D class of materials is its finite band gap unlike in graphene which exhibits zero band gap at the Dirac K-point of the Brillouin zone, and larger charge carrier mobility. These 2D materials are considered to be possible alternatives to graphene and this has invigorated research interest in these materials[6]. Among these 2D materials, metal trichalcogenide titanium trisulfide ($TiS_3$), a direct band gap material ($E_g$~1 eV) with stable layered structure, exhibits in-plane anisotropic geometry[28-34]. As a result, it shows remarkable photo response, gate tunability, robust thermoelectric performance and strong quasi-one dimensional physical properties[32-33]. *Ab-initio* calculations predict an isotropic carrier mobility of a single layer $TiS_3$[32-33] with larger carrier mobility along *b-axis* ($13.87 \times 10^3 cm^2 V^{-1} s^{-1}$) and that along *a-axis* is relatively marginal ($1.01 \times 10^3 cm^2 V^{-1} s^{-1}$). The reported carrier mobility in $TiS_3$ nano-sheet[32] along the "*b*" and "*a*" axes are 80 $cm^2V^{-1}s^{-1}$ and 40 $cm^2V^{-1}s^{-1}$, respectively. Interestingly, $TiS_3$ nanoribbon is found to possess superior optical tunability as compared to nano-sheets[32]. The $TiS_3$ stabilizes in a monoclinic structure with space group $C_{2h}^2$ (*P2₁/m*) with two formula units per unit cell (Z = 2)[31]. The crystal structure of a monolayer $TiS_3$ is depicted in **Fig. 1**. The Ti-S bond length along the *b*-axis (0.245 nm) is shorter than the bond along the *a*-axis (0.265 nm)[34] which results in its in-plane an isotropic geometry. The metal ions Ti are located at the centre of distorted trigonal prisms and these trigonal faces of prism are connected to form a chain along the b-axis (**Fig. 1a**). These covalently bonded parallel chains form a sheet, and such sheets are held together by van der Waals force. Hence the TiS$_3$ crystal can be viewed as a set of stacked parallel sheets with each sheet being composed of 1D chains of triangular TiS$_3$ unit[32]. Two kind of sulfur atoms are seen in the structure of TiS$_3$: bridge sulfur and sulfur-sulfur pair atoms (**Fig. 1b**). Although 2D an isotropic materials such as TiS$_3$ are of great practical importance[35-38] investigations of the effect of pressure on TiS$_3$ materials are still scarce. Structural study of 2D materials subjected to pressure in a diamond anvil cell is an effective way to tune their structure[39] and electronic[40] properties. Besides, Raman spectroscopic studies on such layer materials can be useful to estimate their thermal transport properties[39,40]. Interestingly, the mechanical strain in graphene[41] and MoS$_2$[40] are reported to alter their electronic structure. TiS$_3$ exhibits a strong Raman signal[30,34] as in graphene[39] and MoS$_2$[40]; hence, Raman spectroscopy is expected to give useful insight about its structural along with mechanical stability and thermal transport properties. Raman studies [39,41] on 7-layer graphene nanosheets up to 40 GPa indicate a change in G-band slope and its excessive broadening above 16 GPa is attributed to its deviation from 2D-layer structure. Pressure induced iso-structural electronic transition in multi-layered MoS$_2$ from a semiconducting to a metallic state at 19 GPa has been inferred from high pressure (HP) Raman spectroscopy[40]. Out of plane A$_{1g}$ Raman mode is an intense band that is sensitive to the electronic transition[40]. HP Raman spectroscopic studies on a TiS$_3$ whisker of ~100 μm thickness suggested no phase transition up to 26 GPa[37] with inconclusive phonon modes behaviour at high pressure. Kang *et. al.* using *ab*-initio calculations on electronic band structure of TiS$_3$ film suggests its band gap to be independent of the number of layers, stacking order and vertical strain, and attribute this result to the absence of charge density redistribution in the interlayer region[42]. Recently a strain-induced dynamical instability in a TiS$_3$ monolayer was obtained from phonon dispersion calculations. It was revealed that TiS$_3$ becomes dynamically unstable for an applied uniaxial tensile strain larger than 6% (8%) along the *a- (b)* direction[43] which is yet to be examined experimentally. Anharmonicity of phonon modes is responsible for various thermal properties such as thermal expansion and thermal conductivity of materials. By investigating Raman spectra as a function of pressure and temperature independently, complete information about the quasi-harmonic and true anharmonic components of the phonon modes and the implications can be analysed[44]. In an earlier study[30], we reported our temperature dependent Raman spectroscopic studies on $TiS_3$ nanoribbons in range 88K-570 K. Four out-of-plane $A_g$ phonon modes linearly soften (red-shift) with temperature and the $TiS_3$ compound was found to be stable in this temperature range. However, the pressure dependent phonon vibrational properties of few-layer $TiS_3$ nanoribbons remain unexplored. Therefore, it is of interest to investigate the effect of high pressure on the vibrational properties of $TiS_3$ to identify the anharmonic phonons and consequently its structural instability. More over contribution of these phonons to thermal expansion can also be estimated using high pressure mode behaviours. To our knowledge, there are no reports of the effect of pressure on vibrational properties of $TiS_3$ nanoribbons or estimation of its thermal expansion using the phonon spectroscopic results.

In this work, we report pressure-dependent Raman spectroscopic studies of $TiS_3$ nanoribbons synthesized by direct solid-gas reaction method. TEM studies on the as synthesized materials suggest a high crystalline nature of the as prepared sample. *In-situ* high pressure Raman spectroscopic studies in a diamond anvil cell were carried out up to 34 GPa to study the phonon behavior under compression and any possible phase transition(s) at high pressure. Anomalies in the slope of mode frequency, line-width and their Raman intensity in $TiS_3$ nano-ribbon areattributed to a possible onset of structural transition at 13 GPa, subsequently resulting in a complete structural phase transition at 43 GPa. Pressure coefficients and mode Grüneisen parameters of the $A_g$ modes were obtained from their Raman spectra. Contribution of these phonons to the thermal expansion across the *a-b* plane (along *c*-axis) is estimated using our experimental mode Grüneisen parameters. To obtain reliable information on phase transition and Grüneisen parameter, Raman spectra were measured at close intervals of pressure.

## RESULTS AND DISCUSSION

Fabrication of TiS$_3$ nanoribbons were accomplished by using solid-gas reaction method **(Fig. 2(a-c))**. The details on the synthesis are reported in the experimental section and our earlier studies.[30,32] The typical schematic of Diamond anvil cell used for high pressure Raman measurements is depicted in Fig. 2(d); the close view schematic of the sample chamber is shown in **Fig. 2(e)** to have a better understanding of sample, ruby, pressure transmitting medium and gasket position between two anvils.

**Morphological structure:**

The morphology of as prepared TiS$_3$ materials was investigated using high magnification SEM **(Fig. 3(a, b))**. SEM micrograph shows the nanoribbons of length 30-80 μm and width about 2 μm. The HRTEM image confirms growth of ribbon along a (011) plane with lattice spacing 0.41 nm in accordance with JCPDS data card no. 00-015-0783. AFM image and its height profile **(Fig. 3(c, d))** indicate the thickness of the nanoribbon is about ~30 nm suggesting that each nanoribbon is consists of multi layers of TiS$_3$ (~43 layers).

**High pressure Raman spectroscopy:**

At ambient temperature and pressure the crystal structure (Space group $P2_1/m$) of TiS$_3$ belongs to C$_{2h}$ point group with two formula units per unit cell.[45] Factor group analysis suggests the distribution of Raman active phonons among different irreducible representations[46] as $\Gamma = 8A_g + 4B_g$. The non-vanishing polarizability tensor corresponding to these Raman modes are:[47] $A_g = \begin{bmatrix} a & 0 & d \\ 0 & b & 0 \\ d & 0 & c \end{bmatrix}$, and $B_g = \begin{bmatrix} 0 & e & 0 \\ e & 0 & f \\ 0 & f & 0 \end{bmatrix}$.

The Raman spectrum of TiS$_3$ nanoribbon recorded at ambient temperature (**Fig. 4**) shows four prominent A$_g$ Raman bands located at 175, 300, 370 and 557 cm$^{-1}$, similar to the number of modes reported earlier [30]. These band positions are obtained by fitting the Raman spectrum to a Lorentzian function. These observed Raman active phonon modes vibrate along the *c*-axis, perpendicular to the quasi-1D chain. B$_g$ phonon modes involving atomic displacement along the chain (*b*-axis) are weak[30, 45]; hence they are not discernible in the Raman spectrum. The Raman band at 175 cm$^{-1}$ arises due to rigid vibrations of 1D-TiS$_3$ chain[31,45]. On the other hand, the bands located at 300 (degenerate band) and 370 cm$^{-1}$ correspond to the internal vibrations of polyhedral TiS$_6$ unit, and the degenerate band at 557 cm$^{-1}$ arises due to predominant contribution from non-bridging S-S pair vibration (diatomic vibration) of trigonal prism[31,37]. For clarity, schematic representations of the motion of atoms for different pressure sensitive Raman active phonon modes are represented in **Fig. 1c**. To examine the changes in the Raman spectrum of TiS$_3$ nanoribbon under the influence of high pressure the Raman spectra were recorded in the DAC at different pressures (**Fig. 5**). At 0.6 GPa, the spectral features are found to be similar to the ambient one (**Fig. 4**). As pressure is increased, the intensity of Raman bands except that at 557 cm$^{-1}$ reduces and peaks are broadened. Although the rigid band at 175 cm$^{-1}$ survives at high pressure, its intensity reduces significantly and becomes weak at high pressure. The band intensity of 557 cm$^{-1}$ mode grows as pressure is increased up to ~13 GPa and reduces thereafter up to 34 GPa, the highest pressure of the present studies. In addition, the line-width of this band is found to be insensitive to pressure up to ~13 GPa and soften continuously upon further compression. The band located at 484 cm$^{-1}$, close to the highest frequency S-S diatomic vibrational band, has been attributed to nearly degenerate band at 557 cm$^{-1}$ [37, 45]. Appearance of this band at high pressure (20-25 GPa) can be considered to be due to their different pressure dependencies. The line-width behaviour of Raman bands with pressure is discussed later. At high pressure, as can be seen in **Fig. 5(b)**, the 557 cm$^{-1}$ band shifts to lower frequency. Usually, as can be expected, upon compression, the phonon mode frequencies increase as bond lengths involving atomic vibration are reduced. However, sometimes mode frequency decreases under the influence of pressure and such modes are known as soft modes[44]. Existence of such soft modes often suggests instability of crystal lattice and eventually the system undergoes a structural phase transition. In the present case, the decrease of the 557 cm$^{-1}$ mode frequency with pressure is due to the increase of S-S bond length possibly due to electron-phonon interaction. Often in 2D materials[40], the softening of phonon is due to the effect of electron-phonon interaction owing to the reduction in interlayer spacing under compression. More evidence of phase transition is noticed from the pressure dependencies of mode frequencies and other spectral parameters such as mode line-width and intensity. The spectral parameters are obtained by quantitative analysis of the pressure dependent Raman spectra using Lorentzian function. The pressure evolution of Raman mode frequencies (**Fig. 6**) shows anomalous behaviour of these modes around ~13 GPa that can be attributed to the onset of structural transition resulting from structural distortion. Upon increasing pressure, a complete structural transition is evident as discussed later. Theory of mean field approximation relates soft phonon with structural transition[48]. The square of the soft phonon mode frequency ($\omega^2$) is related to its transition pressure by the relation, $\omega^2 \propto (P - P_c)$, for $P < P_c$, and the soft phonon vanishes as the pressure approaches a transition pressure ($P_c$). The variation of $\omega^2$ as a function of pressure is depicted in **Fig. 7(a)**. As can be noticed, $\omega^2$ falls linearly up to 34 GPa, the highest pressure covered in the present investigation. In the absence of experimental data above 34 GPa, the fitted parameters obtained from the linear curve fitting of the experimental data were used to extrapolate the mode frequency with respect to pressure in the high pressure region (>34 GPa), and found that the 557 cm$^{-1}$ $A_g$ soft mode vanishes at 43 GPa, the expected complete structural transition pressure. It can be mentioned here that first-principles calculations of strain dependence of phonon dispersion curves of TiS$_3$ monolayer predicted a structural instability[43] at large strain values of 6%-8%. Upon increasing strain up to 6%-8%, imaginary vibrational frequencies are observed at Γ-point of the reported phonon dispersion curves pointing towards a structural transition,[43] and that theoretically predicted transition is corroborated experimentally in our present HP Raman spectroscopic investigation. The pressure dependencies of the line-widths of the observed phonon modes are shown in **Fig. 8**. The line-widths of the rigid chain mode at 175 cm$^{-1}$ and internal modes located at 300 and 370 cm$^{-1}$ broaden at high pressure. These internal modes line widths are found to nearly double between 13 and 34 GPa. Upon increasing pressure, the anisotropy of the Ti-S octahedra increases. As a result, the distribution of Ti-S bond lengths is expected to increase. This leads to an increase in the inhomogeneous broadening of these internal vibrations. The broadening of rigid mode and the corresponding decrease in mode intensity with increasing pressure indicate a possibility of growth of static positional disorder[44, 49]. On the other hand, the line-width of highest frequency band at 557 cm$^{-1}$ is found to be insensitive to pressure up to ~ 13 GPa, followed by an increasing trend and thus shows an anomaly around 13 GPa as depicted in **Fig. 8**. The line-width of this mode increases by a factor of four between 13 and 34 GPa. It can be mentioned here that such anomaly in line-width of phonons was observed when materials undergo either a structural[50] or an electronic transition ($MoS_2$[40] and $Sb_2Se_3$[51]).

In order to identify the phase transition in 2D materials, an intensity analysis of Raman mode is useful[40]. This is due to the fact that light couples directly with the polarization in materials and manifests in its intensity[52]. The integrated Raman intensity of the band at 557 cm$^{-1}$ was normalized with respect to that of the band at 300 cm$^{-1}$. The latter band corresponds to the internal vibrations of TiS$_6$ octahedron, which persists over the entire pressure range of the present study and exhibits a slight change in its intensity upon compression. The pressure dependence of the normalized intensity of 557 cm$^{-1}$ mode is shown in **Fig. 7b**. One can notice that the intensity increases up to 13 GPa and subsequently falls beyond this pressure. Hence the observed anomaly in mode intensity around 13 GPa is similar to the onset of transition pressure as noticed from **Fig. 6** and **Fig. 8**, and further corroborates that the $TiS_3$ nanoribbons undergo a pressure-induced structural transition which begins at~13 GPa. The Raman spectra were also recorded in the pressure-released cycle and their peak positions with pressure follow the same trend as observed in the compression cycle is shown in **Fig. 5**. After complete pressure-release, the Raman spectrum is found to be reversible and matches with the ambient spectrum obtained from the pristine $TiS_3$nano-ribbon sample as shown in **Fig. 4**. This suggests that the pressure-induced structural transition of $TiS_3$ nanoribbons is completely reversible.

The pressure coefficients of Raman bands were obtained by linear fit to $\omega_i$ vs. $P$ curves in ambient and HP phases(**Fig. 6**).The low frequency $A_g$ mode which is attributed to the rigid chain vibration has a large pressure dependency, while the 557 cm$^{-1}$ mode is insensitive to pressure (slope ~zero) in the ambient phase. Frequencies of modes at 175, 300 and 370 cm$^{-1}$increase upon compression and follow normal hardening behaviour, whereas the highest frequency mode located at 557 cm$^{-1}$ becomes almost constant in the ambient phase below 13 GPa. The pressure dependencies of modes in the HP phase (13-34 GPa) are found to be less compared to those of the ambient phase. The pressure coefficient of the band at 557 cm$^{-1}$ is found to be negative (**Table 1**). The reduced slopes $\frac{1}{\omega_i}\frac{\partial \omega_i}{\partial P}$ of phonon modes, which essentially represent the bond strength of phonon vibrational modes[44]inboth phases, were obtained (**Table 1**).In the ambient phase, the reduced slope of 175 cm$^{-1}$ rigid chain mode is larger than others indicating that this mode is more sensitive to pressure and gets easily compressed and deformed, whereas a low value of reduced slopes for the 300 and 370 cm$^{-1}$ internal modes suggest that these vibration are comparatively less sensitive to pressure. Since the normalized slopes of modes in HP phase are lower than those in the ambient phase (**Table**

**1**), the mode vibrations in HP phase are less influenced by compression and the bonds involving these atomic vibrations are stiffer. The band at 557 cm$^{-1}$ has a moderate value of negative reduced slope pointing towards its contribution to negative thermal expansion in the HP phase. Mode Grüneisen parameters of A$_g$ modes which exhibit out-of-plane vibration (along *c*-axis), are calculated using the relationship,[44] $\gamma_i = (\omega_i \chi_c)^{-1} \frac{\partial \omega_i}{\partial P}$, where $\chi_c$ is the axial compressibility along the *c*-axis. Using the reported $\chi_c^{-1}$=50GPa[37], $\gamma_i$ values were calculated using our experimental pressure dependencies of mode frequencies. Mode Grüneisen parameters $\gamma_i$ of A$_g$ modes of the ambient phase are listed in Table 1. Larger value of $\gamma_i$ of the rigid mode 175 cm$^{-1}$ implies more contribution to the thermal expansion coefficient. These microscopic parameters $\gamma_i$ are key input for obtaining thermal properties such as specific heat, thermal expansion coefficient of materials, as described below.

Contribution of observed A$_g$ phonon modes to thermal expansion coefficient $\alpha$ of TiS$_3$ nanoribbon can be estimated using $\alpha = (\gamma_{av} C_V)/(3 V_m \chi_c^{-1})$, where C$_v$ is the molar specific heat at constant volume, $\gamma_{av}$ is the average Grüneisen parameter, V$_m$ is the molar volume and $\chi_c^{-1}$ is the axial compressibility along c-axis. Using Einstein's specific heat, $C_i = R[y_i^2 \exp(y_i)]/[\exp(y_i) - 1]^2$, where $y_i = \hbar\omega_i/k_B T$ and R is the gas constant, the total specific heat C$_v$ can be obtained ($i$ = 1 to 4) and $\gamma_{av}$ is defined as $\gamma_{av} = 1/2 (\sum p_i C_i \gamma_i)/C_V$, where p$_i$ is the degeneracy of phonon at Brillouin zone centre. Using the unit cell volume, the molar volume is calculated as 4.439 × 10$^{-5}$ m$^3$/mole; the reported $\chi_c^{-1}$ value is taken as 50 GPa[37], and considering our experimental mode Grüneisen parameters, the linear thermal expansion along the c-axis at 300 K is estimated as $\alpha$ = 1.321 × 10$^{-6}$ K$^{-1}$, which is found to be of the same order of magnitude as the one reported in MoS$_2$[53].

As mentioned earlier, the change in mode frequency due to the effect of temperature has two contributions. One is due to phonon-phonon interaction at constant volume known as intrinsic anharmonic contribution (explicit) and other one is implicit contribution that solely depends on lattice volume change. The latter implicit quasi harmonic contribution is obtained from pressure dependence of mode frequency and enables one to infer the explicit, true anharmonicity of phonon modes. The Raman mode frequencies of TiS$_3$ observed by compression at 0.6 GPa and cooling down to 88 K[30] are listed in **Table 1**. Inspection of these Raman shift values indicates that several of the mode frequencies obtained by way of compression is nearly the same as that obtained by volume contraction achieved by cooling. This suggests that most of the mode frequencies behaviour with temperature is mainly due to quasi harmonic contribution (caused by only change in lattice volume). However, the mode at 370 cm$^{-1}$ has a significant difference in its Raman shift with temperature and pressure indicating highly anharmonic nature which arises from phonon-phonon interaction. The present phonon spectra studies at high pressure provide insight into the role of several A$_g$ phonons in the structural phase transition and thermal expansion in TiS$_3$ nanoribbons. Our experimental finding is consistent with the theoretically predicted structural phase transition from phonon dispersion calculations.

**CONCLUSIONS**

In conclusion, we report pressure dependent Raman spectroscopic studies of out-of-plane A$_g$ phonons in titanium trisulphide nanoribbons, prepared by direct solid-gas reaction method and infer their cross-plane thermal expansion coefficient. Out of four prominent observed A$_g$ modes, the phonon modes at 175, 300, 370 cm$^{-1}$ show blue shift upon compression while that located at 557 cm$^{-1}$ exhibits softening about ~13GPa. Anomalies in phonon mode frequencies and excessive broadening in line-width of the soft phonon about ~ 13 GPa are attributed to the onset of a structural transition. Complete structural phase transition at 43 GPa is observed from extrapolation, consistent with recent reported *ab-initio* theoretical predictions. The first order pressure coefficients of $A_g$ modes are found to lie in the range 0.3-1.5 cm$^{-1}$/GPa. The experimental mode Grüneisen parameter of the 175 cm$^{-1}$ vibrational $A_g$ mode is large and sensitive to pressure. Using these mode Grüneisen parameters, cross-plane thermal expansion at 300 K is estimated to be $1.32 \times 10^{-6}$ K$^{-1}$. The structural changes in TiS$_3$ nanoribbon under compression is recovered in pressure released cycle. The observed results can be crucial in calibration and correction of performance of next generation TiS$_3$-based nano-electronics and optical devices under extreme stress conditions.

## EXPERIMENTAL SECTION

Titanium trisulphide layers have been obtained by direct sulfuration of Ti disc (dia : 10 mm) in a vacuum sealed ampoule at 550 °C for 20 hours (please see the schematic in **Fig. 2(a-c)**). The used Ti discs were etched in HNO$_3$: HF (30:4 wt %) for 1 minute, followed by washed in distilled water, and subsequently dipped in ethanol. Heating rate was 0.4±0.05 °C/min and the cooling was took place in ambient conditions. The detailed synthesis procedure for TiS$_3$ nanoribbons is described in our earlier reports.[30,32] The micro-structure of the materials was investigated using a scanning electron microscopy. Thickness of the nanoribbon was characterized using an Atomic Force Microscope (ICON system, Bruker, Santa Barbara CA, USA) operated in contact mode. *In-situ* high pressure Raman spectroscopic measurements were carried out using a compact, symmetric diamond anvil cell (DAC) with diamonds of culet diameter 500 µm. Raman spectra were recorded using a *micro*-Raman spectrometer (Renishaw inVia) operated in the backscattering configuration. A 514.5 nm line from an Ar-ion laser with an output power of ~0.4mW is used for excitation of the sample. A few flakes of TiS$_3$ sample were loaded into a 200-µm hole of a pre-indented stainless-steel gasket in the DAC. 4:1 methanol-ethanol mixture was used as the pressure transmitting medium and ruby was loaded along with the sample as a pressure calibrant (please see the schematic in **Fig. 2(d-e)**). Measurements were carried out up to ~34 GPa, covering the wavenumber range 100-650 cm$^{-1}$. Raman spectra were also recorded during the pressure-reducing cycle. The pressure transmitting medium is quasi-hydrostatic up to 10 GPa and beyond this pressure the experimental results can be influenced by uniaxial pressure[49,54]. The spectra were analyzed using Lorentzian line shapes using Peak Fit software (JANDEL).

## Author Contributions

D. J. L. and A.C. conceptualized this project. J. O. I., E. F., J. R. A., C. S., I. J. F., H. S. J. Z. A. P., A. C. and D. J. L. prepared and characterized the materials. K. K. M and T. R. R carried out the pressure dependence Raman spectroscopy measurements of TiS$_3$ nanoribbonand processed the data. A. P.and K. K. M. prepared the all schematic representations. All authors analyzed and discussed the data. K. K. M. and D. J. L. wrote the first draft of the manuscript, with all authors commenting and editing the manuscript.

## Competing interests:

The authors declare no competing financial interests.

## Acknowledgement

Dr D J Late thank science and engineering research board (SERB) Government of India for funding support. K.K.M. would like to thank Prof. R. S. Katiyar, University of Puerto Rico (USA) for encouragement. A. P. would like to acknowledge grant (CZ.02.2.69/0.0/0.0/16_027/0008355) for postdoc funding.

**Table 1:** Pressure coefficients and Grüneisen parameters of Raman modes in the ambient phase and high-pressure phase. Numbers in the parentheses are the standard errors in the least significant digit.

| Ambient phase | | | | | | HP phase | |
|---|---|---|---|---|---|---|---|
| Mode frequency(cm$^{-1}$) | | | | | | | |
| 0.6 GPa | 300 K | 88 K[a] | $P$-coefficient (cm$^{-1}$/GPa) | Reduced $P$-coefficient (GPa$^{-1}$) | Grüneisen parameter $\gamma_i$ | $P$-coefficient (cm$^{-1}$/GPa) | Reduced $P$-coefficient (GPa$^{-1}$) |
| 178 (A$_g$) | 175 | 177 | 4.75(4) | 0.027142 | 1.3571 | 3.12(8) | 0.017828 |
| 302 (A$_g$) | 300 | 302 | 2.21(9) | 0.007366 | 0.3683 | 1.07(7) | 0.003566 |
| 374 (A$_g$) | 370 | 371 | 2.93(7) | 0.007918 | 0.3959 | 1.78(1) | 0.004811 |
| 559 (A$_g$) | 557 | 560 | 0[b] | - | - | -0.91(3) | -0.001634 |

[a]Ref. 30

[b]Error larger than slope value is considered as zero

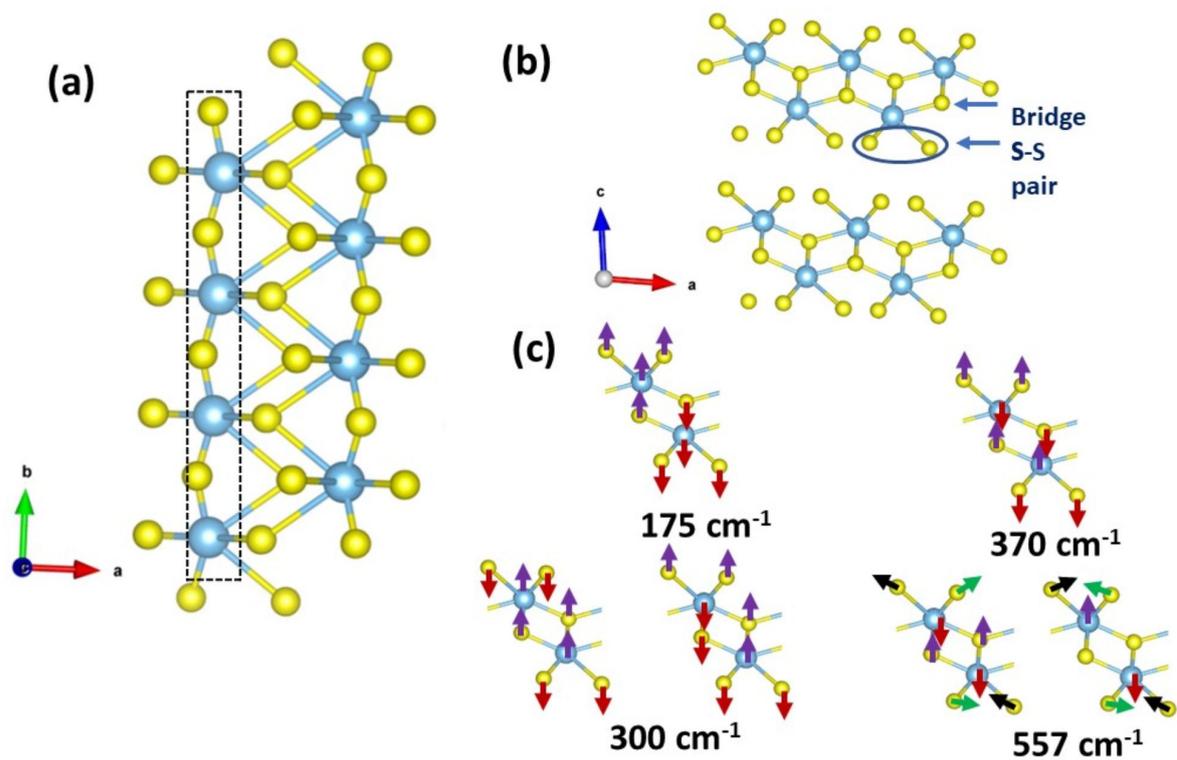

**Figure 1:** Crystal structure of Titanium Trisulfide. Central Ti metal atoms (light blue color) are covalently bonded to six S atoms (yellow color) forming trigonal prism. These prisms share triangular faces to form 1D-chains (shown by dashed lines). The black dashed lines mark the Ti chain along crystallographic b-axis. (b) cross-sectional view of the structure of $TiS_3$, with bridge sulfur and sulfur-sulfur pair atoms. (c) schematic representation of the motion of atoms for different pressure sensitive Raman active phonon modes. The bands at 300 and 557 cm$^{-1}$ are degenerate. Structure models are made using VESTA[55].

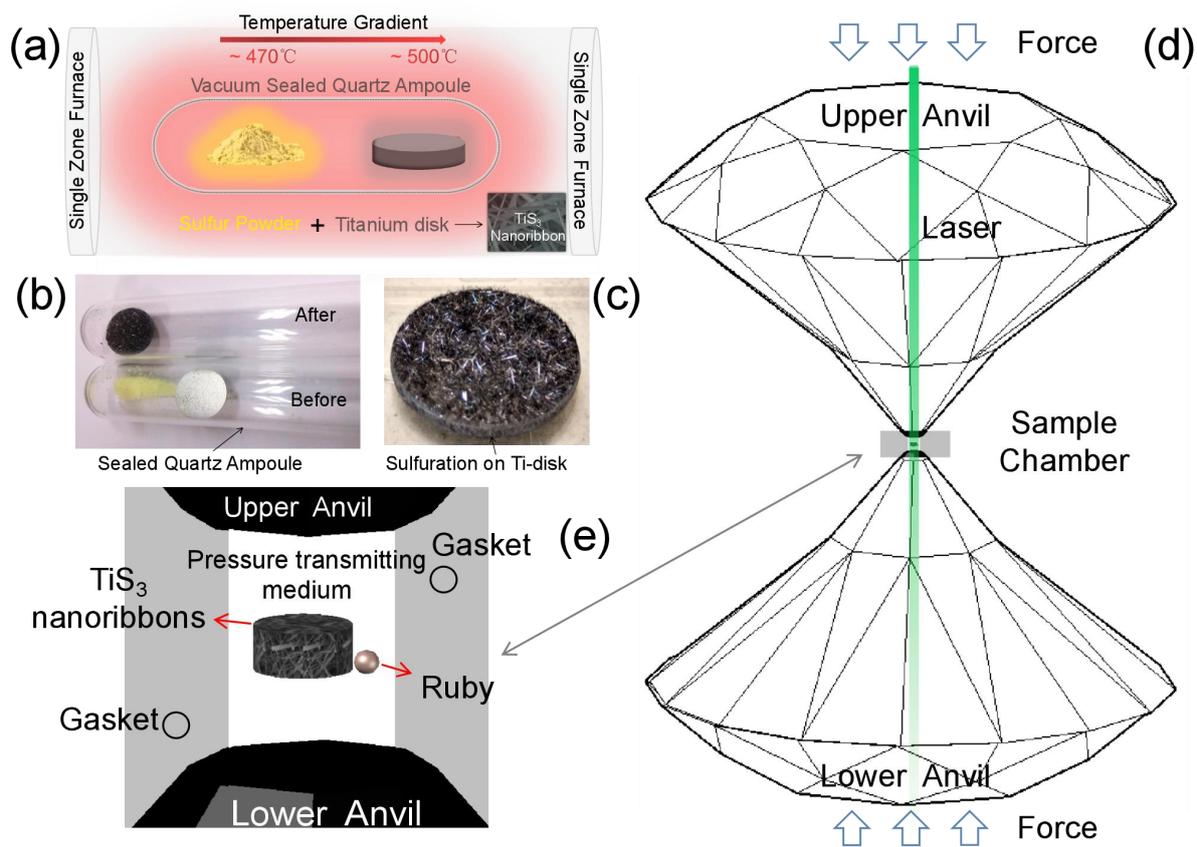

**Figure 2:** (a) schematic of solid-gas reaction method involved in fabrication of TiS$_3$ nanoribbons. For clarity, optical images of (b) sealed ampoules before and after the synthesis process and (c) a zoomed view of sulfurated Titanium disk are shown. A Schematic on (d) diamond and (e) gasket used in the diamond anvil cell are also shown. The schematics are made using AutoCAD software, version 2020.

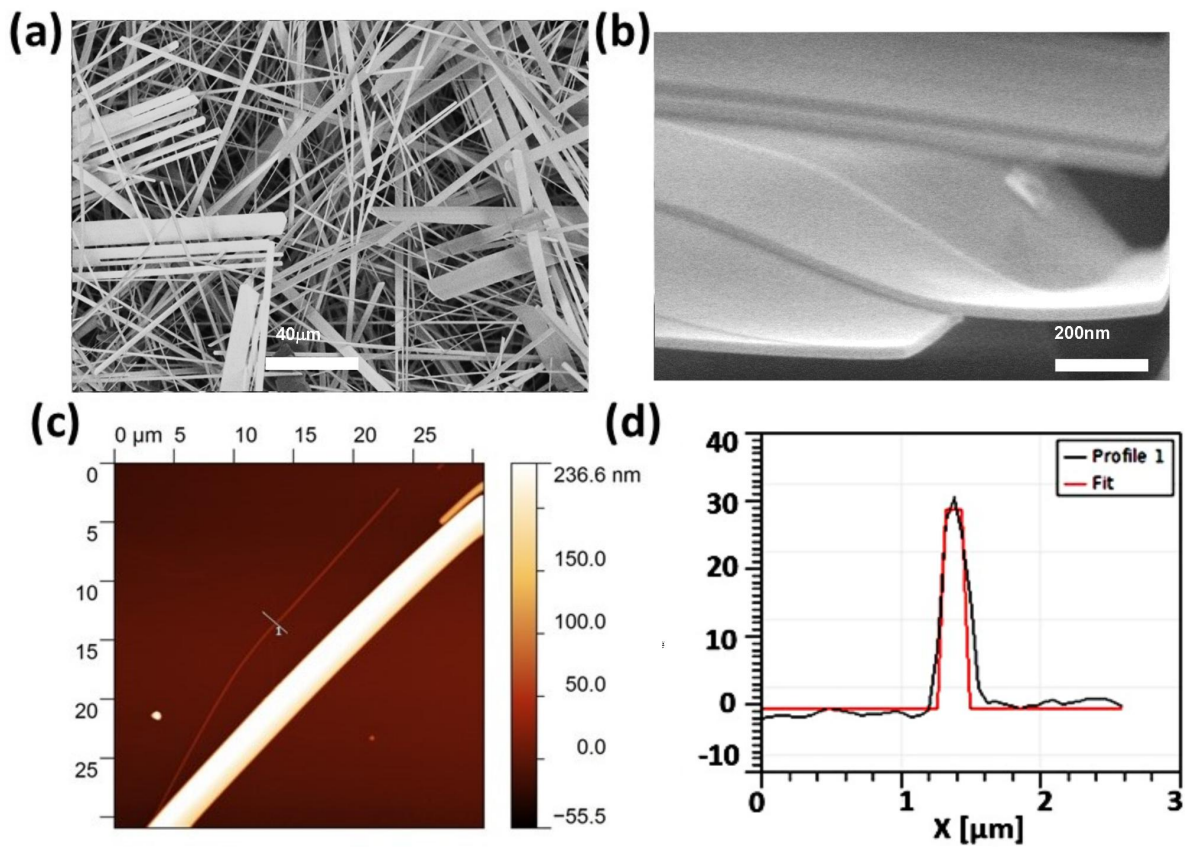

**Figure 3:** Scanning electron microscopy results on TiS$_3$ nanoribbon (a) high magnification SEM image recorded from TiS$_3$ nanoribbon (b) high magnified SEM image of the TiS$_3$ nanoribbon. (c) AFM image and (d) AFM height profile of TiS$_3$ nanoribbon.

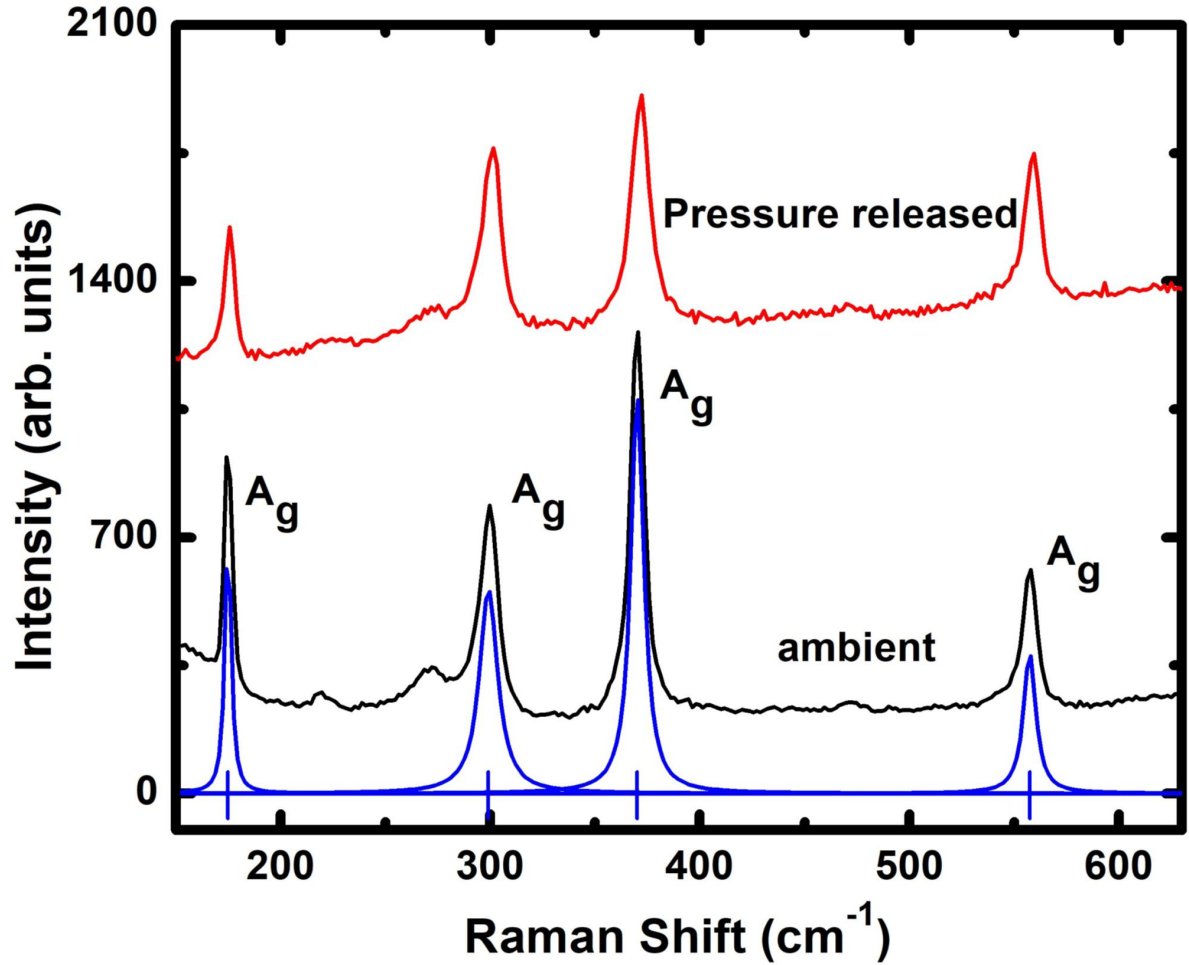

**Figure 4:** Raman spectra of TiS$_3$ nanoribbons at ambient and after complete pressure released in diamond anvil cell. The deconvoluted individual Raman bands (blue colour) are also shown. Blue tick marks locate the calculated prominent Raman bands of the ambient phase. For clarity, the spectra are vertically shifted.

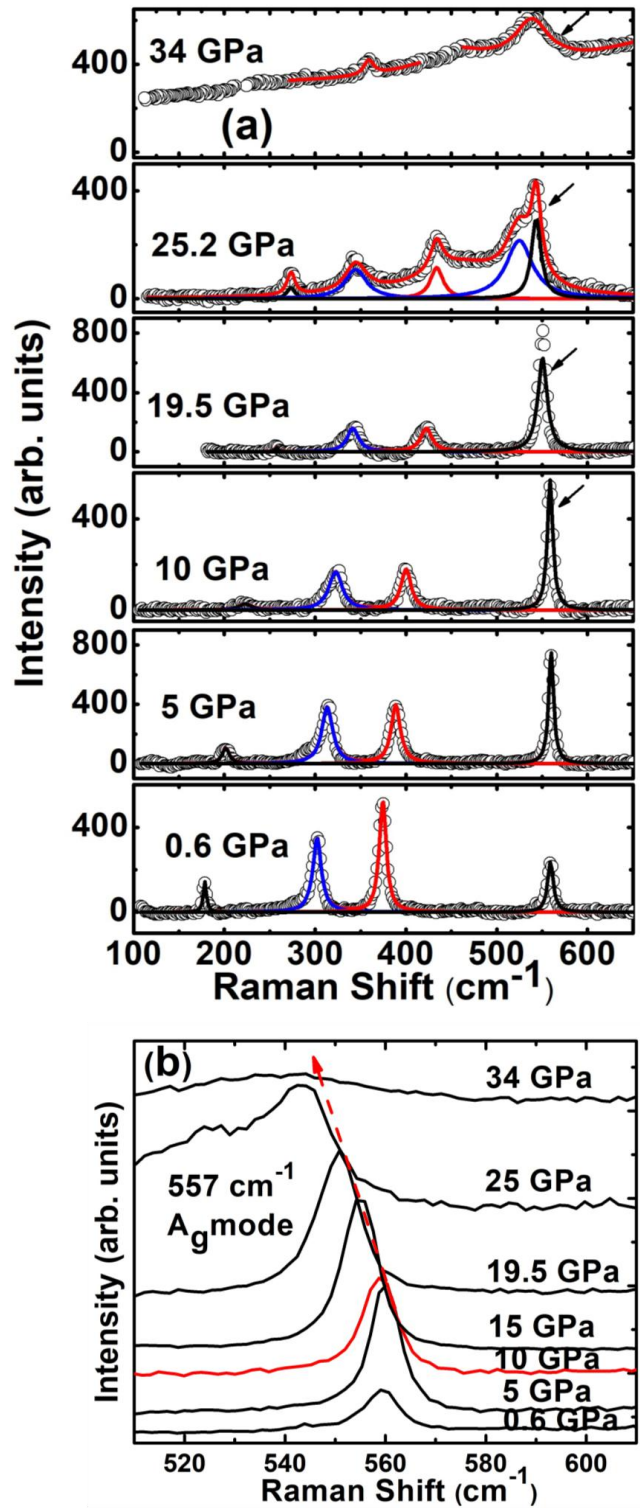

**Figure 5:** Raman spectra of TiS$_3$ nanoribbons at different pressures. Open circles are the experimental data. Solid lines are the fits for individual bands and total fit to the data. Softening of the mode is indicated by arrows, (b) for brevity, A$_g$ mode at 557 cm$^{-1}$ continues to soften upon increase in pressure above 13 GPa is also shown.

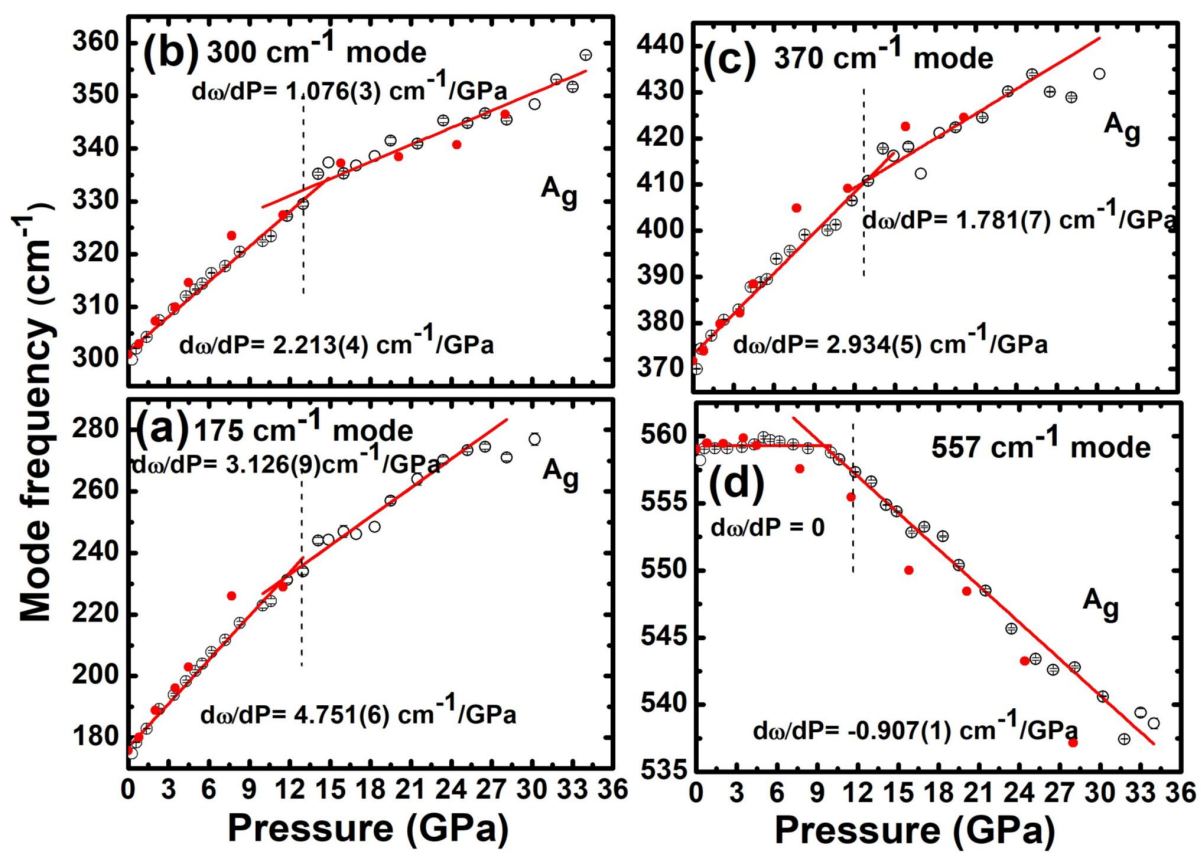

**Figure 6:** Pressure-dependent Raman mode frequencies of TiS$_3$ nanoribbons. Open and solid symbols correspond to data during compression and decompression cycle, respectively. Solid lines are linear fits to the frequencies. A$_g$ mode at 557 cm$^{-1}$ softens and undergoes a dramatic change in slope d$\omega$/d$P$ near 13 GPa. All other A$_g$ modes harden with pressure and exhibit slope change ~13GPa, the onset of transition pressure.

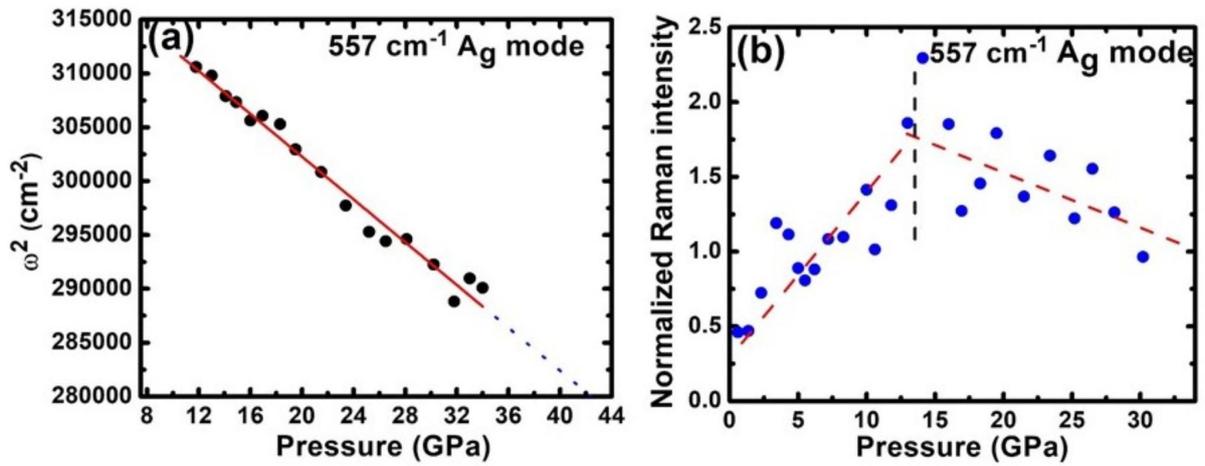

**Figure 7:**(a) Pressure dependence of the square of the Raman-active $A_g$ soft mode frequency. Solid line is a linear fit to the data. Dotted line corresponds to extrapolated line into the high-pressure field indicating the complete structural phase transition at 43 GPa. (b)Pressure dependencies of normalized integrated intensity of 557 cm$^{-1}$ mode with respect to that of mode at 300 cm$^{-1}$. The curve drawn through the data point is guide to the eye. Vertical line shows the anomaly in intensity behaviour around onset of transition pressure ~13 GPa.

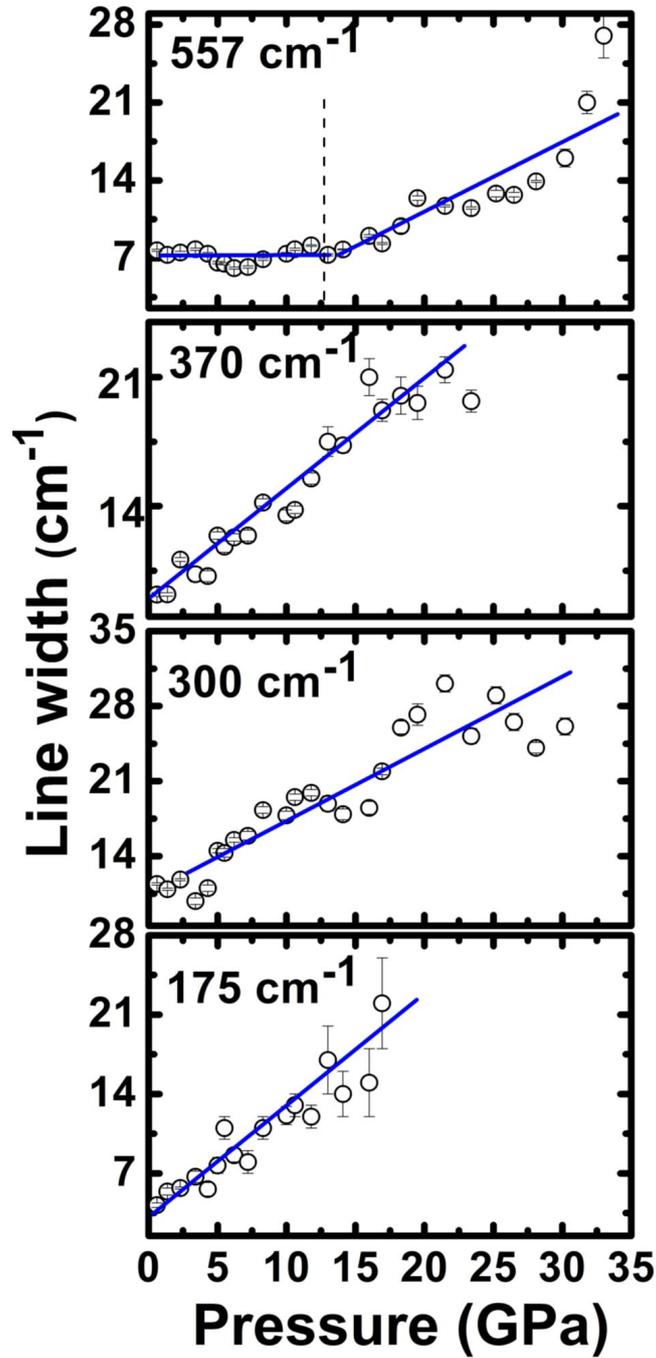

**Figure 8:** Pressure-dependent line-widths of Raman bands of TiS$_3$ nanoribbons. Open symbols correspond to the evolution of band line-widths during compression. Solid lines are guide to the eye. Vertical dotted lines show change of slope at the onset of transition pressure.